\documentclass[a4paper]{aa}

\usepackage{psfig}
\usepackage{times}

\def\msun {\hbox{M$_{\odot}$}} 
\def\msun {\hbox{M$_{\odot}$}}

\begin{document}
   \thesaurus{ 09.09.1,09.03.2, 09.19.2; 13.25.4}

    \title{A comparison of the X-ray line and continuum morphology of Cassiopeia~A}

    \author{Jacco Vink\inst{1} \and
	M. Concetta Maccarone\inst{2}  \and
	Jelle S. Kaastra\inst{1} \and 
	Teresa Mineo\inst{2} \and
	Johan A. M. Bleeker\inst{1} \and
	Andrea Preite-Martinez\inst{3} \and 
	Hans Bloemen\inst{1}
}

    \offprints{J. Vink (J.Vink@sron.nl)}

    \institute{
SRON Laboratory for Space Research, Sorbonnelaan 2, NL-3584
CA Utrecht, The Netherlands 
\and
Istituto di Fisica Cosmica ed Applicazioni all'Informatica, CNR, 
Via U. La Malfa 153, I-90146, 
Palermo, Italy
\and
Istituto di Astrofisica Spaziale,
Area di Ricerca di Tor Vergata,
Via del Fosso del Cavaliere 00133,
Roma}

\date{Received  22 October 1998/Accepted }

\maketitle

\begin{abstract}
We present deconvolved narrow band images of Cas A as observed by the MECS 
instrument on the BeppoSAX X-ray satellite. 
The images show that Cas A has different morphologies in the continuum and
line bands. This difference points towards a synchrotron origin for part of the
X-ray continuum of Cas A.
Going to the hardest continuum band we find that the emission is
coming predominantly from the Western region, making this the most
likely location for the hard X-ray tail seen with instruments on 
CGRO (OSSE), BeppoSAX (PDS) and RXTE (HEXTE).
\end{abstract}

\section{Introduction}
The X-ray emission from supernova remnants has generally been understood to be 
thermal emission (i.e. brems\-strahlung and line emission) for the so called 
shell type supernova remnants, such as Tycho, Kepler and Cassiopeia~A (Cas A), 
and  synchrotron emission for the plerions such as the Crab nebula. 
In the latter case the synchrotron emission is caused by cosmic ray electrons 
accelerated in the magnetosphere surrounding the neutron star. 
The ideas on shell type remnants have, however, changed dramatically with the
publication of the analysis of ASCA data of the remnant of SN1006
by Koyama et al. (1995). 
They proposed that the featureless X-ray spectra of the rims of
the remnant were the result of synchrotron emission (earlier anticipated by 
Reynolds \& Chevalier \cite{Reynolds81}).
SN1006 is a shell type remnant and does not contain, 
as far as we know, a pulsar, hence the synchrotron emission should arise from
cosmic ray electrons accelerated in the blast wave of the supernova remnant.
This view is supported by the recent discovery of TeV gamma rays coming from 
the Northeastern rim of SN1006 (Tanimori et al. \cite{Tanimori}).  

Of course, if there is X-ray synchrotron radiation coming from SN1006, then it 
is very likely that other (young) shell type remnants also produce synchrotron 
radiation at some level. Although, it is worth pointing out that SN 1006 was
probably a type Ia supernova, whereas Cas A is very likely the remnant of
a core collapse supernova (Type Ib or II). 
Recently, a hard X-ray tail was discovered in the spectrum of Cas A, 
the youngest known galactic supernova remnant (The et al. \cite{The}, Favata et al. \cite{Favata},
Allen et al. \cite{Allen}). 
This tail can be attributed to synchrotron radiation, 
but it could also arise from bremsstrahlung caused by a non-Maxwellian tail 
to the thermal electron distribution. 
Note that, like SN1006, Cas A does not seem to contain a pulsar.
In the case of Cas A there is no doubt that a substantial part of the 
emission is thermal in nature, since the remnant displays strong line emission. 
So, whatever the nature of the hard tail, at lower photon energies this 
component is intermingled with thermal components.

From a theoretical point of view both synchrotron emission 
(Reynolds \cite{Reynolds98}) and
non-thermal bremsstrahlung can exist (Asvarov et al. \cite{Asvarov}). 
For instance, the cosmic ray electrons responsible for the radio emission of
Cas A find their origin in thermal electrons accelerated at the shock. 
This means that at all energies intermediate between the thermal electron
distribution and the relativistic electrons responsible for the radio emission,
a population of electrons should exist. 
The Coulomb equilibration process is slow, 
so if at some moment the initial  electron distribution is non-thermal, 
the relaxation towards a Maxwellian distribution is likely to evolve slowly
(Laming \cite{Laming98}), because
the relaxation time scales with electron energy ($E$) as $E^{-3/2}$.
To give an example applicable to Cas A ($n_{\rm e} \sim 10$~cm$^{-3}$): 
at 1~keV the relaxation time is $\sim 1$~yr, 
at 10~keV it is $\sim 30$~yr, and at 50~keV its is of 
the order of the age of the remnant.
Therefore, from say 20~keV to 1~GeV the electron distribution should 
be a continuously
decreasing function and non-thermal bremsstrahlung should exist, but may be
less pronounced than the synchrotron emission.

So, what can we say at present about the relative contributions of a 
synchrotron component and a non-thermal brems\-strahlung component to the hard X-ray tail?
Ultimately this will be resolved by observing X-ray polarization, 
which is not feasible presently.
Here we attempt to bring the discussion somewhat further by 
investigating the spatial distribution of the X-ray emission up to 10.5~keV.
We do this by presenting deconvolved images in narrow bands obtained by the
MECS instrument on board BeppoSAX.

Before discussing our analysis, a short characterization of Cas A.
The remnant is at a distance of $3.4^{+0.3}_{-0.1}$~kpc 
(Reed et al. \cite{Reed}) and
is between 340 and 318 years old. The younger age is based on the
possible explosion date of 1680, when Flamsteed observed an unidentified 6th 
magnitude star near the current position of Cas A (Ashworth \cite{Ashworth}).
The older age is based on the expansion analysis of optical filaments 
(van den Bergh \& Kamper \cite{vdBergh83}).
The current wisdom is that the progenitor was probably an early type
Wolf-Rayet star (Fesen et al. \cite{Fesen87}) with an initial mass of about
30\msun\ (Garci\'a-Segura et al. \cite{GLM96}) and a final mass 
less than 10\msun\ (Vink et al. \cite{VKB96}). 
Recent expansion studies of high resolution X-ray images indicate that the 
current overall expansion velocity is $\sim$ 3200~km/s for the bright shell 
and $\sim$ 5200~km/s for the fainter emission associated with the blast wave
(Vink et al. \cite{Vink98}, Koralesky et al. \cite{Koralesky}).
This is considerably greater than the expansion of the bright radio ring
which is about 2000~km/s (Anderson \& Rudnick \cite{AR95}).

\begin{figure}[htbp]
	\psfig{figure=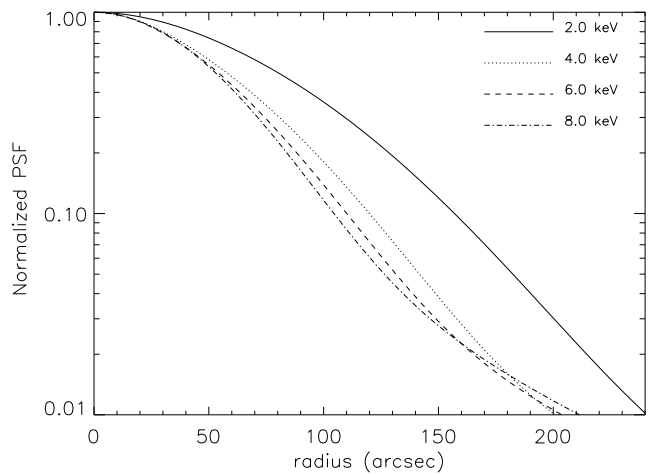,width=8.8cm,angle=90}
	\caption[]{The MECS3 point spread function at four energies: 2~keV, 4~keV, 6~keV and 8~keV.\label{psf}}
\end{figure}

\section{Observations and analysis}
\subsection{The data}
The BeppoSAX satellite (Boella et al. \cite{Boella97a})
contains four X-ray detectors:
the LECS (0.1--10~keV) and MECS (1.5--12~keV), composed of three units,
the HPGSPC (4--120~keV) and the PDS (13--300~keV). The overall spectrum of 
Cas A as observed by BeppoSAX is shown in the paper by Favata et al. (\cite{Favata}).
Only the LECS and MECS have imaging 
capabilities. Here we have opted for only using the MECS
(Boella et al. \cite{Boella97b}) instrument,
since the supporting structure of the LECS blocks part of the field of view,
whereas the MECS has an unobstructed view of Cas A, which simplifies the
image analysis considerably.
Furthermore, the MECS has a higher sensitivity above 5~keV than the LECS.
BeppoSAX observed Cas A five times; four observations during the performance 
verification phase in August and September 1996 and one observation on
November 26, 1997.
The total observation time is different for each set of instruments, 
but data presented here comprise 161~ks of MECS observation time. 
The data analyzed here are based on the MECS2 and MECS3 units
(labeled together as MECS23) for uniformity:
the MECS1 was no longer functioning at the time of the last observation.
Comparing the various data sets we found that there were rather large
inaccuracies in the aspect solutions. 
This was most distinct for the last 
observation ($\sim 1.4$\arcmin) made after 
a gyroscope failure, which may have caused the relatively large
error in the aspect solution.
Using a correlation analysis similar to that
employed by Vink et al. (\cite{Vink98}), we shifted the observations 
in order to match in position. 

\begin{table}
        \caption{The energy ranges and number of iterations for the 
Lucy-Richardson deconvolution used for the images. 
Images dominated by line emission are labeled with the element producing the
line emission. In all cases the emission consists of blends of lines of
H-like and He-like ions.
\label{lucy}}
	        \begin{flushleft}          
		\begin{tabular}{lll}
			\hline\noalign{\smallskip}
characterization & energy range &\# iterations \\
& keV &\\
			\hline\noalign{\smallskip}    
Si        & 1.6--2.4  & 440 \\
S         & 2.4--2.9  & 244 \\
Ar        & 2.9--3.3  & 158 \\
continuum & 4.0--6.0  & 117 \\
Fe K      & 6.1--7.1  & 114 \\
continuum & 7.3--9.0  &  64 \\
continuum & 9.0--10.6 &  43 \\
                	\hline\noalign{\smallskip}    
		\end{tabular}
        \end{flushleft}
\end{table}

\begin{figure}[ht]
	\psfig{figure=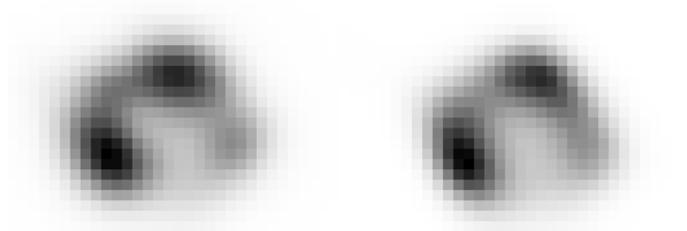,width=8.8cm,angle=-90}
	\caption{ A comparison of the MECS23 image in the silicon band 
(on the right) 
with the ROSAT PSPC image from $\sim 1.5$~keV to $\sim 2.3$~keV convolved with
a gaussian with $\sigma = 24$\arcsec.\label{pspc}}
\end{figure}

\subsection{Lucy-Richardson deconvolution}
The MECS point spread function (PSF) consists of two components, representing
the gaussian shaped intrinsic resolution of the MECS instrument and the 
scattering component originating from the mirrors.
These components can be identified in the following parameterization of the 
PSF as a function of radius, $r$, (Boella et al. \cite{Boella97b}):
\begin{equation}
f(r) = C \{\alpha \exp( -\frac{1}{2} \frac{r^2}{\sigma^2} ) + (1.0 + \frac{r^2}{R^2} )^{-\beta}\};
\end{equation}
note that the parameters $\sigma$ , $R$, $\alpha$ and $\beta$ 
are energy dependent (S. Molendi, private communication). 
$C$ is a normalization constant, so that the integration over the total 
plane equals 1.
Fig.~\ref{psf} shows the PSF of the MECS3 for four different energies. 
As can be seen the PSF changes with energy, especially the change from 
2 to 4 keV is dramatic. Above 4~keV the PSF does not change much anymore.
At higher energies the resolution of the MECS increases,
but the mirror scattering becomes worse.
Fortunately, the photon statistics at 2~keV are far superior to those
at 8~keV, so
we can compensate for the degraded resolution at 2~keV by performing more
iterations of the deconvolution algorithm.

We used the analytical representation of the PSF to deconvolve the 
narrow band MECS images with the Lucy-Richardson deconvolution algorithm 
(Lucy \cite{Lucy}, Richardson \cite{Richardson}).
One problem associated with this algorithm is the number of iterations that 
should be used. Too many iterations may cause spurious results, 
too few iterations do not bring out the full potential of the image. 
We solved this by statistically comparing the convolved model image with 
the raw image at each iterative step; 
when the model improvements were less than $1\sigma$ we 
halted the deconvolution process.
The consistency of our results was verified by comparing the deconvolved 
images of individual datasets with those of the total dataset and, 
as discussed below, with the results from other X-ray telescopes.
We point out that some of the features presented below were
already recognizable in the raw images (Maccarone et al. ~\cite{Maccarone}).

\begin{figure}[h]
	\psfig{figure=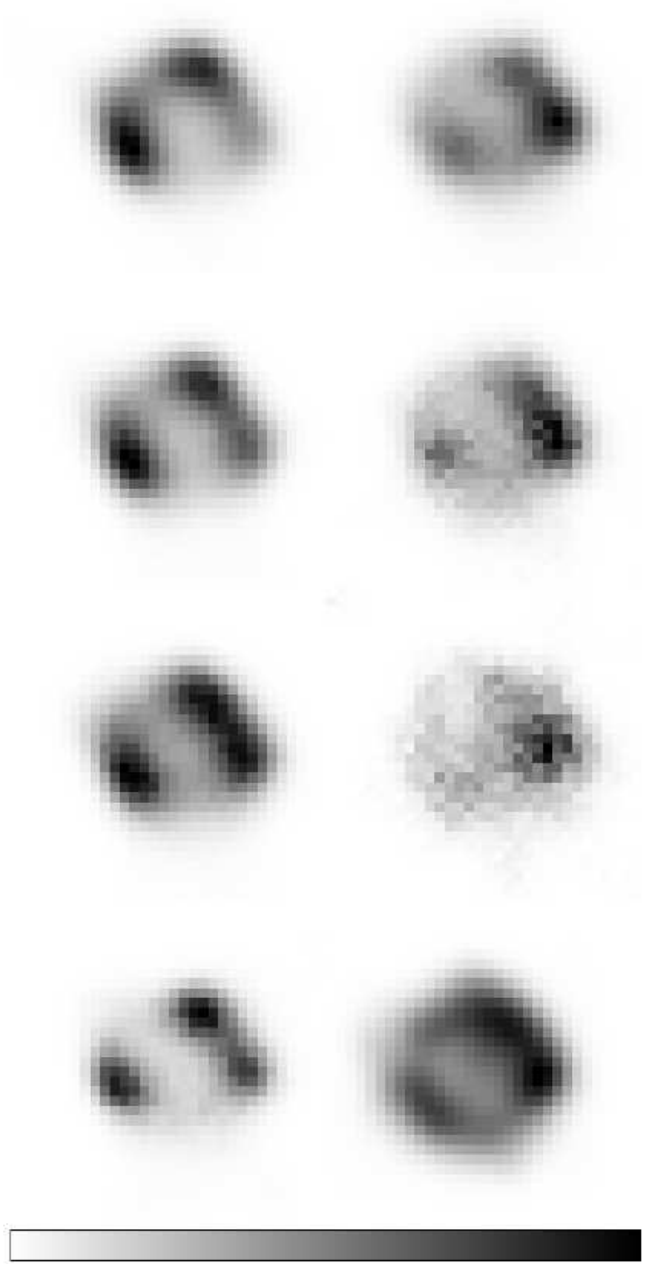,width=8.8cm}
	\caption{ The deconvolved MECS23 images. 
On the left: image dominated by line emission from Si, S, Ar and Fe. 
On the right: 
continuum emission in the energy ranges 4-6~keV, 7.3--9~keV, 9-10.6~keV and a VLA image convolved with a Gaussian with $\sigma = 24$\arcsec.
Note that the line images are not corrected for the continuum distribution,
which will affect the Ar and Fe K image most.
\label{mecs23}}
\end{figure}

\subsection{Comparison with previous measurements}
We present here the first analysis of the morphology of Cas~A for 
photon energies in excess of 6~keV.
A similar analysis was done with ASCA SIS0 data for energies below 6~keV
(Holt et al. \cite{Holt}). The advantage of the MECS instrument over ASCA SIS
is that it is more sensitive above 6~keV and also the PSF of 
the MECS is simpler. The core of the MECS PSF is, however, broader than for 
the ASCA SIS, so that the deconvolved MECS images in the range 1.8~keV 
to 6~keV do not show as much detail as the deconvolved SIS images.
Apart from the lack of detail, 
the MECS images below 6~keV are in agreement with
the images presented by Holt et al. (\cite{Holt}).

A comparison of the ROSAT PSPC image with the deconvolved silicon image
provides another check on the validity of our results. As can be seen in
Fig.~\ref{pspc} the images compare well.

\begin{figure}[ht]
	\psfig{figure=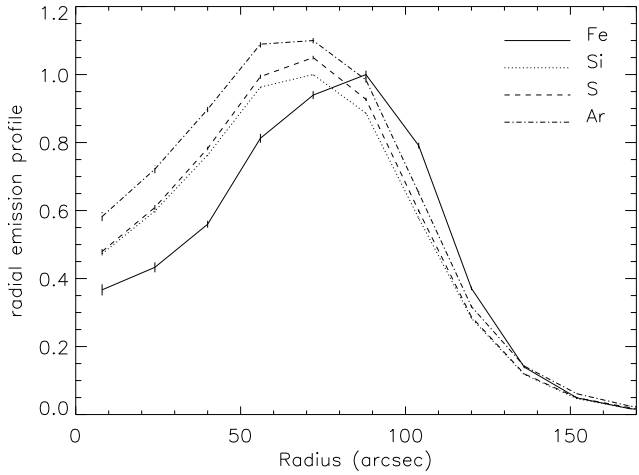,width=8.8cm,angle=90,clip=2}
	\caption{The radial profiles for the Southeastern region (45\degr to 170\degr from the North).
The iron emission peaks about 20\arcsec\ further from the center than the 
emission from Si, S and Ar.\label{profile}}
\end{figure}

\section{Results}
The deconvolved narrow band images are displayed in 
Fig.~\ref{mecs23}; the energy ranges and the number
of Lucy-Richardson iterations are listed in Table~\ref{lucy}.
In Fig.~\ref{mecs23} we have ordered the images such that images dominated by
line emission are on the left side and continuum images are on the right.
For comparison we included an archival VLA image showing 
synchrotron emission at a wavelength of 21.7~cm. 
This image has been smoothed to roughly the same resolution as the deconvolved
MECS images.

\begin{figure}[ht]
\begin{flushleft}
	\psfig{figure=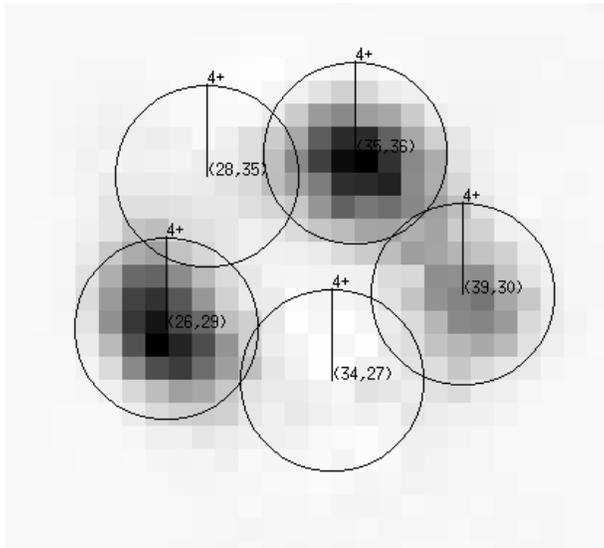,width=8.0cm}
\end{flushleft}
	\caption{ This image (the deconvolved, continuum subtracted iron K 
image) shows the regions
used for producing Fig.~\ref{spec}. North is up. 
\label{regions}}
\end{figure}

It is immediately clear that the spatial distribution
of the continuum emission is different from the line emission, whereas the
images relating to the line emission do not differ much from one another.
The silicon image is not so bright in the West, 
but this is due to the fact that 
this part of the remnant is more absorbed by the interstellar medium than 
the rest of Cas A (Keohane et al.~\cite{Keohane98}). 
There is also a hint in the Fe K band image that the iron line 
emission in the Southeast peaks further out from the center of the remnant 
than the line emission of Si, S and Ar. The displacement is about 20\arcsec, as can be seen in Fig.~\ref{profile}.
This may indicate that the iron emission
in that area is predominantly coming from the shocked circumstellar medium
rather than from the shocked ejecta.
The lack of variation from one line image to the other 
indicates that dust depletion cannot play a major role in explaining
the line radiation morphology, 
since silicon and iron can be easily dust depleted, whereas
sulphur and argon cannot.

\begin{figure}[ht]
\begin{flushleft}
	\psfig{figure=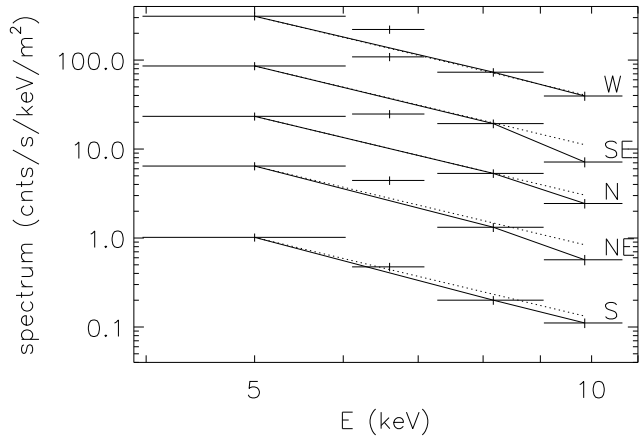,width=8.0cm}
\end{flushleft}
	\caption{ Spectra in the energy range 4.0--10.6~keV reconstructed from 
the deconvolved images. The different spectra are for circular regions with
radii of 64\arcsec. The vertical error bars represent 5\% errors,
the horizontal bar shows the energy ranges.
The solid lines indicate the assumed continuum emission.
The dotted lines show spectra with a power law index of 3.
The spectra have been rescaled for reasons of clarity.\label{spec}}
\end{figure}

The continuum emission does vary strongly from one energy band 
to another, but the continuum images differ from the line emission images 
in that they all peak in the Western region of the remnant,
a property they share with the VLA image.

\begin{table}
        \caption{The results of the analysis of the continuum and Fe K band 
images (cf. Fig.~\ref{spec}). The errors include 5\% systematic errors
for the flux measurements. $\Gamma$ is the power law index. 
The index @ 6.6~keV is based on the 4-6~keV and 7.3-9~keV images 
and the index @ 9.0~keV is based on the 7.3-9~keV and
9-10.6~keV images.\label{spec_tab}}
	        \begin{flushleft}          
		\begin{tabular}{llll}
			\hline\noalign{\smallskip}
region & EW Fe K$\alpha$ & $\Gamma$ @ 6.6 keV &  $\Gamma$ @ 9.0 keV \\
       & keV &\\
			\hline\noalign{\smallskip}    
Northeast & $0.69 \pm 0.12$ & $3.2 \pm 0.2$ & $4.5 \pm 0.6$ \\
North     & $1.45 \pm 0.18$ & $3.0 \pm 0.2$ & $4.2 \pm 0.5$ \\
West      & $0.61 \pm 0.12$ & $3.0 \pm 0.2$ & $3.3 \pm 0.5$ \\
South     & $0.17 \pm 0.09$ & $3.3 \pm 0.2$ & $3.2 \pm 0.5$ \\
Southeast & $1.94 \pm 0.22$ & $3.0 \pm 0.2$ & $5.4 \pm 0.5$ \\
                	\hline\noalign{\smallskip}    
		\end{tabular}
        \end{flushleft}
\end{table}

\begin{figure}
	\psfig{figure=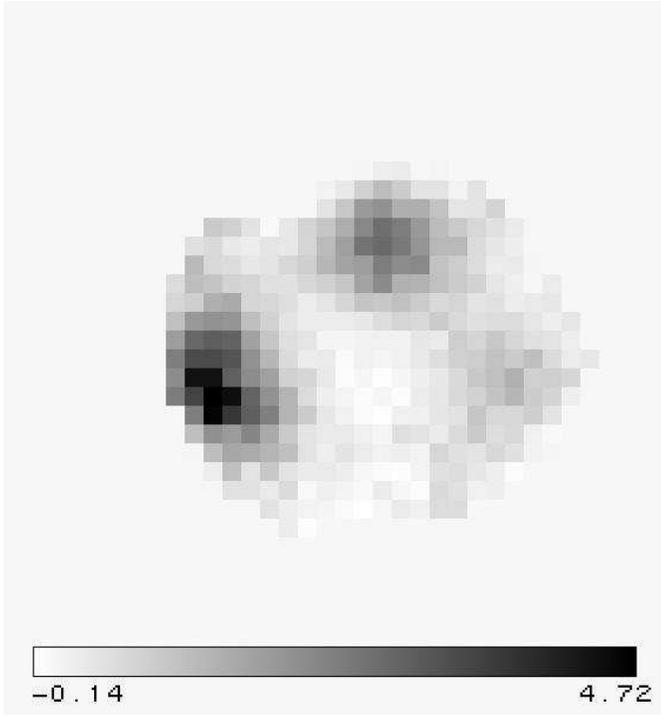,width=8.8cm}
	\caption{This image represents the iron K-shell emission equivalent 
width (in keV). Negative values, which are unphysical, 
have been removed at the edge of the remnant. Pixels are 16\arcsec\ in  size.\label{fe_ew}}
\end{figure}

In order to make more quantitative statements we concentrate on the emission
above 4~keV. Each image was corrected for the bandwidth and instrument
effective area, so that the pixel values correspond to the approximate 
flux density per pixel (i.e. photons keV$^{-1}$m$^{-2}$s$^{-1}$pixel$^{-1}$, 
one pixel being 16\arcsec\ in size).
With these images rough spectra at various regions were made (see Fig.~\ref{regions} for the regions).
This is shown in Fig.~\ref{spec} and useful parameters related to the 
spectra are listed in Table~\ref{spec_tab}. The statistical errors of the 
flux measurements are small, so the systematic errors, 
which are of the order of 5\%, dominate.
We corrected the iron image for the continuum contribution. Dividing this
image by the estimated continuum emission at 6.6~keV we ended up 
with an image approximating the equivalent width of the iron K-shell 
emission (Fig.~\ref{fe_ew}). 
The peak equivalent width in Fig.~\ref{fe_ew} of 4.7~keV is rather high, 
but the average equivalent width per region, as determined from the images, 
agrees with the value in Table~\ref{spec_tab} and spectral analysis results.

The continuum in the iron band image was estimated by interpolating 
the 4--6 keV and 7.3--9.0~keV image to 6.6~keV assuming a photon 
index of 3.0 and averaging those two continuum images.
The results, shown in Fig.~\ref{spec}, Fig.~\ref{fe_ew} 
and Table~\ref{spec_tab},
confirm that the line emission peaks in the North and Southeast, whereas
the continuum peaks in the West. 
The Western regions becomes relatively brighter at higher energies.

If line emission is an indication for a thermal origin of 
the emission, is its prominent absence a signature for synchrotron emission?
It seems at least the most natural explanation, since other explanations
involving steep abundance gradients or ano\-malous electron distributions are 
more far fetched. 
This argument is supported by the observation that all the line images 
have rather similar morphologies, 
so if there are abundance gradients they should be roughly equal 
for all elements.
Non-thermal bremsstrahlung does not explain the difference 
between the continuum morphology and the line morphology, 
because the electrons involved in this process should also cause line 
emission.
The line emission does indicate, however, that there is also a
contribution of bremsstrahlung to the continuum. 
The bending of the spectra in some regions  shown in Fig.~\ref{spec}
(e.g. the Southeastern and Northeastern regions) indicates that the continua
in those regions are probably dominated by thermal-bremsstrahlung, 
whereas the Western and Southern regions are likely to be dominated by 
synchrotron emission.
The trends visible in the continuum images make it very likely that the 
hard X-ray tail above 20~keV 
(The et al. \cite{The}, Favata et al. \cite{Favata},
Allen et al. \cite{Allen}) is coming predominantly from the West of Cas A.

Models for the X-ray synchrotron emission of shell type supernova remnants 
(e.g. Reynolds~\cite{Reynolds98}) predict that the synchrotron spectrum 
extrapolated from the radio emission should cut off with a term shallower
than $\exp(-(\frac{E}{E_c})^\frac{1}{2})$, with $E$ the photon energy and
$E_c$ the cut off energy. 
From this we can derive that the photon index, $\Gamma$, 
of the X-ray spectrum is given by:
\begin{equation}
\Gamma = \alpha_R + \frac{1}{2}(\frac{E}{E_c})^\frac{1}{2} + 1,
\end{equation}
with $\alpha_R$ the radio spectral index, which is 0.78 for Cas A.
For thermal bremsstrahlung the photon index is:
\begin{equation}
\Gamma = 1 + \frac{E}{kT_{\rm e}},
\end{equation}
with $kT_{\rm e}$ the electron temperature.
So we see that the overall value of $\Gamma = 3$ at 6.6~keV 
(cf. Holt et al.~\cite{Holt}) implies $E_c = 1.1$~keV or 
$kT_{\rm e} = 3.3$~keV (in agreement with Vink et al. \cite{VKB96}).
The thermal spectrum steepens much faster than the synchrotron 
spectrum: at 9.0~keV the synchrotron model predicts $\Gamma = 3.2$,
whereas the bremsstrahlung model predicts $\Gamma = 3.8$.
So we see that some spectra fall off too rapidly to be dominated by 
synchrotron emission, except the spectra of the Western and Southern region.

It has been argued (Keohane et al. \cite{Keohane96}) that in the Western region
the blast wave is interacting with the molecular cloud seen in 
absorption towards
Cas A (Bieging et al. \cite{Bieging}).
Such an event could lead to an enhanced cosmic ray production,
but this hypothesis implies that the thermal emission 
should peak in the Western region as well
since there should also be more shock heated material. 
The line images, which trace the thermal component, 
do not support this.
Because the infrared image of Cas A (Lagage et al. \cite{Lagage}) 
does not peak in the Western region, depletion in dust cannot explain the
relative lack of emission.
In addition, sulphur and argon are not depleted in dust.
However, 
the lack of line emission can be reconciled with an interaction 
between a molecular cloud and Cas A, if the electron temperature is too low
to produce appreciable S, Ar and Fe K line emission in this region. 
The electron temperature is potentially low if the shock wave decelerates
rapidly when entering the dense cloud.
The weak line emission originating from the Western region may 
in that case come from the far side of Cas A, 
where the blast wave is possibly undisturbed.

Turning the attention now to the line images, we note several interesting features.
First of all, little line emission originates from the Southern region.
Although the VLA image also shows some lack of emission in 
that region, the near absence of line emission is rather puzzling.
Another feature is that the morphology of the line images indicates that the remnant can be divided into roughly two halves,
with the intersection running from Northeast to Southwest. 
Interestingly, this is also the division made on the basis of X-ray 
Doppler shift maps (Markert et al. \cite{Markert}, Holt et al. \cite{Holt} and
Tsunemi et al. \cite{Tsunemi}). 
Tsunemi et al. (\cite{Tsunemi}) report that the Fe K line has larger Doppler
shifts than the other elements, although there are some uncertainties, 
due the fact that non-equilibrium ionization effects may mimic Doppler shifts.
If Doppler shifts are indeed larger for the Fe K lines, 
then this may be connected to the observation presented here that the Fe K 
emission in the Southeast peaks further out from the center than the emission 
from other elements. 
In this case most Fe K emission in the Southeast presumably originates
from the 
swept up circumstellar medium, which has a larger velocity than the supernova
ejecta heated by the reverse shock.

\section{Conclusion}
We have presented narrow band images of Cas A deconvolved with the 
Lucy-Richardson method. There is a clear distinction between the morphology 
of the continuum emission and the line emission. 
The images dominated by line emission resemble each other, 
but we found that the iron K-shell emission in the Southeast peaks 
$\sim $20\arcsec\ further out from the center than the emission from other elements.

The continuum emission increasingly enhances in the Western region 
when going to harder X-rays, making it likely that the 
hard X-ray tail is coming predominantly from the Western side of the remnant.
Furthermore, the difference in morphology between the line and continuum
images indicates that part of the continuum emission of especially the Western 
region is probably synchrotron radiation.

The enhanced non-thermal emission from the Western region may be indicative of
a collision of Cas A with a molecular cloud. This hypothesis does not seem
to be supported by the line emission images, which should also 
show brightness enhancement in the West. 
However, a low electron temperature of the 
shocked molecular cloud material may circumvent this.

Our analysis provides the first images of Cas A at energies above 7~keV, which
allows for the separation of the non-thermal and the thermal X-ray emission
at angular scales of the order of 1\arcmin. 
In the near future XMM, with its high throughput above 8~keV, 
will allow extension of this type of research to the arcsecond scale.
The excellent photon statistics expected from this X-ray observatory 
will allow accurate Doppler shift measurements and will potentially
reveal new details, which may provide an explanation to 
some of the peculiarities of the X-ray emission from the Western region.

\begin{acknowledgements}
We thank Silvano Molendi for providing us with the parameterization of the 
MECS PSF. 
M.C. Maccarone and T. Mineo acknowledge useful discussions with Bruno Sacco.
Jacco Vink acknowledges pleasant and useful discussions with Glenn Allen.
We thank the referee J. Dickel for some suggestions
which helped to improve this article.
This research has made use of data obtained through the High Energy
Astrophysics Science Archive Research Center Online Service, provided
by the NASA/Goddard Space Flight Center and 
by the NCSA Astronomical Digital Image Library (ADIL).
This work was financially supported by NWO,
the Netherlands Organization for Scientific Research.
\end{acknowledgements}

\end{document}